\documentclass[preprint]{revtex4-2}
\usepackage{amsmath}
\usepackage{amssymb}
\usepackage{graphicx}% Include figure files
\usepackage{dcolumn}% Align table columns on decimal point
\usepackage{bm}% bold math
\usepackage{xcolor}
\usepackage{mathtools}
\usepackage{braket}
\usepackage{siunitx}

\newcommand{\chiint}{\chi_\text{int}}

\begin{document}

%\preprint{APS/123-QED}

\section*{Title}
Ultrafast Coherent Bandgap Modulation Probed by Parametric Nonlinear Optics

\section*{Author list}
Sebastian Klimmer$^{1,2}$, Thomas Lettau$^3$, Laura Valencia Molina$^{1,2}$, Daniil Kartashov$^{4,5}$, Ulf Peschel$^{3,5}$, Jan Wilhelm$^{6,7}$, Dragomir Neshev$^{2}$ and Giancarlo Soavi$^{1,5,\star}$

\section*{Affiliations}
\noindent
$^1$Institute of Solid State Physics, Friedrich Schiller University Jena, Helmholtzweg 5, 07743 Jena, Germany
\newline
$^2$ARC Centre of Excellence for Transformative Meta-Optical Systems, Department of Electronic Materials Engineering, Research School of Physics, The Australian National University, Canberra, ACT, 2601, Australia
\newline
$^3$Institute of Condensed Matter Theory and Optics, Friedrich Schiller University Jena, Max-Wien-Platz 1, 07743 Jena, Germany
\newline
$^4$Institute of Optics and Quantum Electronics, Friedrich Schiller University Jena, Max-Wien-Platz 1, 07743 Jena, Germany
\newline
$^5$Abbe Center of Photonics, Friedrich Schiller University Jena, Albert-Einstein-Straße 6, 07745 Jena, Germany
\newline
$^6$Regensburg Center for Ultrafast Nanoscopy (RUN), University of Regensburg, 93040 Regensburg, Germany
\newline
$^7$Institute of Theoretical Physics, University of Regensburg, 93053 Regensburg, Germany
\newline

$^{\star}$ giancarlo.soavi@uni-jena.de

%\keywords{nonlinear optics, SHG, valleytronics, 2d materials}
\maketitle

\section{Abstract}
Light-matter interactions in crystals are powerful tools that seamlessly allow both functionalities of sizeable bandgap modulation and non-invasive spectroscopy. While we often assume that the border between the two regimes of modulation and detection is sharp and well-defined, there are experiments where the boundaries fade. The study of these transition regions allows us to identify the real potentials and inherent limitations of the most commonly used optical spectroscopy techniques. Here, we measure and explain the co-existence between bandgap modulation and non-invasive spectroscopy in the case of resonant perturbative nonlinear optics in an atomically thin direct gap semiconductor. We report a clear deviation from the typical quadratic power scaling of second-harmonic generation near an exciton resonance, and we explain this unusual result based on all-optical modulation driven by the intensity-dependent optical Stark and Bloch-Siegert shifts in the $\pm$K valleys of the Brillouin zone. Our experimental results are corroborated by analytical and numerical analysis based on the semiconductor Bloch equations, from which we extract the resonant transition dipole moments and dephasing times of the used sample. These findings redefine the meaning of perturbative nonlinear optics by revealing how coherent light-matter interactions can modify the band structure of a crystal, even in the weak-field regime. Furthermore, our results strengthen the understanding of ultrafast all-optical control of electronic states in two-dimensional materials, with potential applications in valleytronics, Floquet engineering, and light-wave electronics.

\section{Main text}
\subsection{Introduction}

An essential criterion of modern science and of our comprehension of nature is that any observation of a physical event perturbs the event itself. While this idea was initially postulated almost one hundred years ago in the framework of quantum mechanics~\cite{heisenberg1927anschaulichen,bohr1928quantum}, the principle remains valid even if we perform classical experiments to study the intrinsic properties of a material. For this reason, understanding to which extent our experimental tools can really probe the equilibrium state of a sample without perturbing it has both fundamental and technological relevance. Light-matter interactions are, in this context, the clearest example of this duality: light is one of the most powerful tools to investigate matter in its various phases and forms. At the same time, light provides an invaluable knob to tune its electronic, optical, and thermal properties. For instance, all-optical modulation underlies some of the most fascinating fields of research in contemporary solid state physics, including light-wave electronics \cite{heide2024petahertz}, photo-induced superconductivity \cite{budden2021evidence} and chirality \cite{zeng2025photo}, Floquet engineering \cite{wang2013observation,weitz2024lightwave}, and coherent bandgap modulation \textit{via} the Optical Stark (OS) and Bloch-Siegert (BS) effects \cite{sie2018large,herrmann2025nonlinear}. On the other hand, the use of light as a non-invasive probe is rooted in the history of modern science, with applications ranging from optical microscopy to absorption, Raman spectroscopy, and nonlinear optics \cite{maiuri2019ultrafast,shree2021guide}. 

For nonlinear optics (NLO) in solid state samples, this interplay between detection and modulation is so crucial to even allow the definition of different areas of research. \textit{Perturbative} NLO \cite{boyd2020nonlinear} is based on the assumption that the interacting light (electric field) is sufficiently weak to be treated as a small perturbation to the equilibrium of the crystal. This type of light-matter interaction is described by a Taylor expansion of the material polarization:
\begin{align}
    \boldsymbol{P} = \varepsilon_0\left(\boldsymbol{\chi}^{(1)}\otimes\boldsymbol{E}+\boldsymbol{\chi}^{(2)}\otimes\boldsymbol{E}\boldsymbol{E}+\boldsymbol{\chi}^{(3)}\otimes\boldsymbol{E}\boldsymbol{E}\boldsymbol{E}+...\right),
\end{align}
where the nonlinear susceptibility $\boldsymbol{\chi}^{(n)}$ is linked to the symmetry and Berry curvature of the system at equilibrium \cite{soavi2025signature}. For this reason, perturbative NLO is among the most versatile spectroscopy tools to study crystal symmetries \cite{li2013probing}, magnetic order \cite{fiebig2023nonlinear}, optical resonances~\cite{wang2015giant}, defects \cite{cunha2020second} or strain \cite{mennel2019second} in crystals. In contrast, \textit{strong-field} NLO operates in a regime where electrons can tunnel from the valence to the conduction band of a crystal, and subsequently, high harmonics are generated by intraband currents and/or interband polarization \cite{ghimire2019high}. The fingerprint that defines the cross-over from the perturbative to the strong-field NLO in high harmonic generation (HHG) is typically the power law scaling of the HHG intensity $I(n \omega) \propto I (\omega) ^{\xi}$, described by the scaling factor $\xi$. In the perturbative regime, the harmonic order $n = \xi$, whereas $n > \xi$ in strong-field HHG \cite{zograf2022high}. Here, we show an intriguing exception to this general rule, whereas second harmonic generation (SHG) in the perturbative regime deviates from the power law, $\xi = 2$ in the vicinity of an optical transition. We demonstrate that this behavior arises from the all-optical modulation of the bandgap triggered by the fundamental beam (FB) \textit{via} the OS and BS shifts. Our work redefines our understanding of all-optical coherent bandgap modulation and of the intrinsic limitations of perturbative NLO due to the dual role of light, which simultaneously induces and probes the perturbation of the energy bands in a solid. Furthermore, our findings are comprehensively captured by a theoretical model based on the Semiconductor Bloch Equations (SBEs)~\cite{SchmittRink1988,Lindberg1988,Wilhelm2020,Yue2022}. The obtained analytical~\cite{Aversa1995} and numerical solutions enable us to further retrieve from experiments the fundamental parameters of dephasing time and transition dipole moment at optical resonances.

\begin{figure}[tb]
    \centering
    \includegraphics{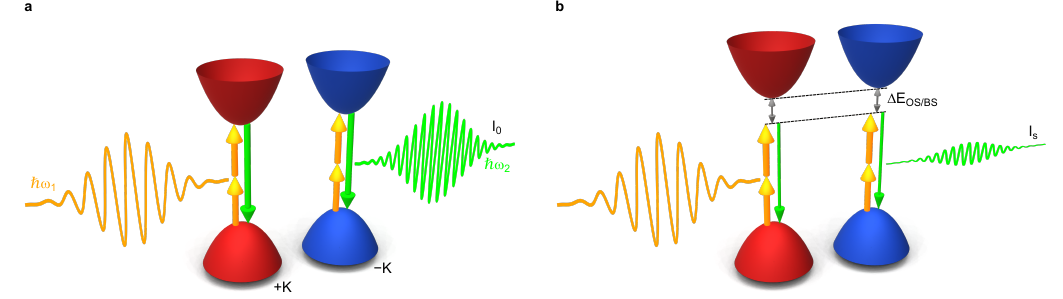}
    \caption{\label{fig:1}\textbf{Principle of ultrafast all-optical bandgap modulation and detection.} \textbf{a}, Sketch of resonant SHG in the $\pm$K valleys of a TMD without considering the effect of all-optical bandgap modulation by OS and BS shifts. Here, a linearly polarized pulse with photon energy $\hbar\omega_1$ (orange) causes a second-harmonic signal (green) with photon energy $\hbar\omega_2 = 2\hbar\omega_1$ and intensity $I_0$. \textbf{b}, By including the intensity-dependent OS and BS shifts, the $\pm$K valleys undergo a blueshift $\Delta E _{OS/BS}$, reducing the efficiency of the SHG process, and decreasing the SH intensity to $I_S < I_0$.}
    \end{figure}

\begin{figure}[tbh!]
    \centering
    \includegraphics{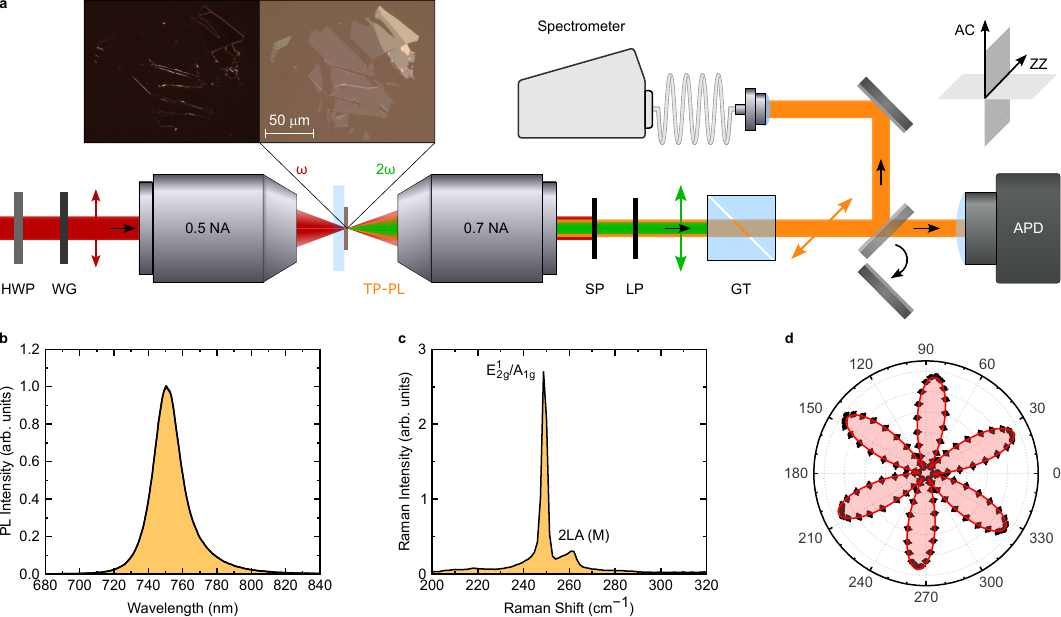}
    \caption{\label{fig:2}\textbf{Experimental setup and sample characterization.} \textbf{a}, The polarization of the tunable pump laser can be defined by a combination of half-waveplate (HWP) and a wire-grid polarizer (WG). The laser is subsequently focused on the exfoliated WSe$_2$ monolayer and the emitted NLO signal is collimated in transmission geometry. Further, a combination of shortpass (SP) and longpass (LP) filters are used for spectral filtering of the signal. A second stage of polarization filtering is done with a Glen-Thompson prism (GT). In a last step, the remaining signal is guided either to a spectrometer or to a silicon avalanche photodiode (APD). The inset shows the bright- and dark-field images of the WSe$_2$ sample. \textbf{b}, PL spectrum of the exfoliated WSe$_2$ monolayer (black). The spectrum shows a pronounced peak centered around $\sim$\;\qty{750}{\nano\meter}, indicating its monolayer nature. \textbf{c}, Raman spectrum showing the degenerate {E$^{1}_{2g}$} and {A$_{1g}$} modes as well as the second-order {2LA(M)} mode. Their frequency difference of $\sim$\;\qty{11}{\centi\per\meter} and the absence of the {B$^{1}_{2g}$} peak at $\sim$\;\qty{309}{\centi\per\meter} underpins further the identification of a monolayer. \textbf{d}, Co-polarized polarization-dependent SHG pattern, measuring the emitted SH signal parallel to the polarization of the incident pump. Black dots represent the experimental data for a fundamental wavelength of \qty{1500}{\nano\meter} and the red line is a $\cos^2$-fit of the data.}
\end{figure}

As a platform for our study, we use a monolayer sample from the family of transition metal dichalcogenides (TMDs), which exhibit several unique features that make them interesting for applications based on parametric nonlinear optics (NLO) \cite{dogadov2022parametric}. These include a large refractive index \cite{nauman2021tunable}, strongly bound excitons that dominate their optical properties even at room temperature \cite{wang2018colloquium}, and a large nonlinear second order susceptibility coefficient combined with the absence of phase matching constraints \cite{Klimmer2021all-optical, trovatello2021optical,trovatello2024tunable,trovatello2025quasi}. In parallel, TMDs provide an excellent playground for all-optical bandgap modulation and valleytronics \cite{sie2018large,herrmann2023nonlinear,conway2023effects,zhoucavity2024,gucci2024ultrafast}. Here, we detect light-induced band structure modulation in a WSe$_2$ monolayer based on parametric SHG. By filtering out the competing signals from two-photon photoluminescence (TP-PL), we isolate the second-harmonic (SH) response of the sample, and we measure an anomalous SH intensity dependence that we can directly relate to the observation of an emerging blueshift of the optical bandgap caused by contributions from both OS and BS shifts (Fig.\;\ref{fig:1}). To achieve this, we conducted wavelength- and polarization-dependent NLO experiments (Fig.\;\ref{fig:2}a; see Methods for details on the experimental setup) across the ${A\!\!:\!\!1s}$ exciton resonance of the WSe$_2$ monolayer sample, focusing in particular on SHG, third harmonic generation (THG) and TP-PL. 

First, we characterized the sample by photoluminescence (PL) (Fig.\;\ref{fig:2}b) and Raman spectroscopy (Fig.\;\ref{fig:2}c) to confirm the monolayer nature of the exfoliated flake (see Methods for details). Polarization optics were then calibrated to enable selective excitation and detection along either the armchair (AC) or zig-zag (ZZ) directions of the monolayer crystal. Furthermore, we characterized the sample with polarization-resolved SHG measurements \cite{rosa2018characterization} at a FB wavelength of \qty{1500}{\nano\meter} (Fig.\;\ref{fig:2}d). The observed symmetric pattern indicates that the sample is not affected by strain, which could otherwise influence the efficiency of nonlinear processes along different crystal directions \cite{mennel2019second}. 

\subsection{Exploiting crystal symmetry for signal filtering}

The aforementioned relaxed phase-matching conditions, resulting from the deep sub-wavelength thickness, are a blessing and a curse at the same time. Whereas TMDs and related heterostructures allow for several NLO process to occur simultaneously \cite{lafeta2021second,hernandez2021nonlinear}, the detailed study of a single process becomes tedious once emission spectra are overlapping. A prominent example for this case is two-photon resonant SHG and TP-PL, which overlap spectrally and thus make any detailed analysis of the SH signal impossible. Since power-dependent SHG at the exciton resonance is the key ingredient for the findings of this work, we isolate the SH signal from the TP-PL using the crystal symmetry of TMD monolayers. When time-reversal symmetry is preserved \cite{herrmann2025nonlinear}, the NLO response of TMD monolayers is defined by the D$_{3h}$ point group, where the second-order susceptibility tensor has only one independent non-vanishing element \cite{boyd2020nonlinear}:
\begin{align}
\chi^{(2)}_{yyy}=-\chi^{(2)}_{yxx}=-\chi^{(2)}_{xxy}=-\chi^{(2)}_{xyx},
\label{eq:1}
\end{align}
where $x(y)$ refers to the zig-zag(armchair) axis of the crystal, respectively. This produces a well-defined polarization dependence of the SH intensity, which in turn can be used for efficient filtering of the TP-PL. In particular, the excitation of a TMD with a pulse polarized linearly along the AC direction results in an SH signal with the same polarization, whereas a pure emission along the ZZ direction can only be achieved by an excitation polarization rotated by 45$^\circ$ with respect to the AC/ZZ axes \cite{Klimmer2021all-optical}. Subsequently, a polarization-resolved measurement with an excitation along AC and with an analyzer along ZZ allows the complete suppression of the SH signal. On the other hand, TP-PL is unpolarized at room temperature \cite{herrmann2023nonlinear}, and it can thus be directly filtered and measured in the aforementioned configuration.

\begin{figure}[tbh!]
    \centering
        \includegraphics{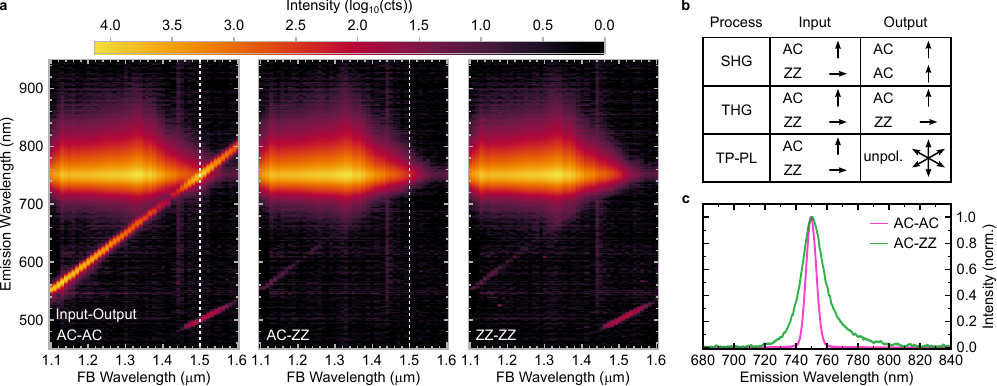}
    \caption{\label{fig:3}\textbf{Crystal symmetry-resolved nonlinear optics in monolayer WSe${_2}$} \textbf{a}, 2D color maps for different combinations of excitation (input) and detection (output) polarization. Parametric wavelength-dependent NLO processes, like SHG and THG, show distinct polarization dependencies, whereas the non-parametric TP-PL (centered at an emission wavelength of $\sim$\;\qty{750}{\nano\meter}) shows an unpolarized emission in all cases. \textbf{b}, Overview of observed connections of input and output polarizations for different NLO processes. \textbf{c}, Normalized intensity of the acquired spectra for the AC-AC (pink) and the AC-ZZ case (green) for a FB at \qty{1.5}{\micro\meter} (white dashed line cut through color maps in \textbf{a}). The much broader FWHM for detection along ZZ indicates a complete suppression of the SH contribution.}
\end{figure}

Thus, having characterized the crystal axes with the measurements in Fig.\;\ref{fig:2}d, we obtain the emission spectra for FB wavelengths ranging from \qtyrange{1.1}{1.6}{\micro\meter} and for various combinations of excitation and detection polarizations with respect to the AC and ZZ axis (Fig.\;\ref{fig:3}a). As expected, we observe distinct emission directions for the simultaneously occurring NLO effects of SHG, THG, and TP-PL, according to the selection rules listed in Fig.\;\ref{fig:3}b. For SHG and THG, the polarization direction of the emitted signal can be directly inferred from the indices of the corresponding NLO susceptibility tensor element (see equation\;(\ref{eq:1}) for SHG and \textit{e.g.}, Ref. \cite{saynatjoki2017ultra} for THG). In contrast, incoherent TP-PL exhibits unpolarized emission, with equal intensities along the AC and ZZ directions. To assess the quality of the signal extinction and thus of our polarization-filtering capability, we compared the emission line shape for an excitation wavelength of \qty{1.5}{\micro\meter}, approximately corresponding to the two-photon resonance of the ${A\!\!:\!\!1s}$ exciton state. At these conditions, the SH signal is strongly enhanced \cite{wang2015giant} as it overlaps with the TP-PL emission. In the AC-AC geometry, the full-width at half-maximum (FWHM) of the SH emission peak, centered at $\sim$\;\qty{750}{\nano\meter}, was determined to be $\Delta\lambda_{SHG}\;=$\;\qty{7.6}{\nano\meter}. This value is consistent with the expected spectral line-width reduction by a factor of $2\sqrt{2}$ \cite{herrmann2023nonlinear} relative to the FB pulse-bandwidth ($\Delta\lambda_\text{FB}=$\;\qty{23}{\nano\meter}, see Ref. \cite{valencia2024enhanced}), confirming that SHG dominates the emitted signal. In contrast, the emission spectrum observed in the AC-ZZ geometry closely resembled the PL spectrum (Fig.\;\ref{fig:2}b), indicating that the TP-PL signal dominates, while the SH signal is effectively suppressed by the Glan-Thompson prism-polarizer in the experimental setup. Thus, by subtracting the AC-ZZ signal (representing the TP-PL contribution) from the AC-AC signal, we could properly filter the SH signal from the background TP-PL signal.

At this point, it is worth commenting on the observation of the TP-PL signal when the photon energy of the FB is exactly half of the 1s exciton energy. Two-photon absorption (TPA) on the 1s exciton state is, in principle, forbidden in the electric dipole approximation \cite{lin2020fast}, and thus our results should be explained either in the context of higher-order effects (electric quadrupole or magnetic dipole \cite{wang2015giant}) or as the effect of 1s-2p exciton state mixing \cite{glazov2017intrinsic}. Similar TPA and TP-PL for excitation (close) to the 1s resonance was reported for a MoSe$_2$ sample embedded in a cavity \cite{lundt2019optical}. Our results confirm the possibility to directly excite the 1s state of excitons in TMDs \textit{via} TPA, an approach that could be used to realize a high degree of valley polarization \cite{wang2013observation}. 

\subsection{All-Optical Bandgap Modulation}

\begin{figure}[h!]
    \centering
    \includegraphics{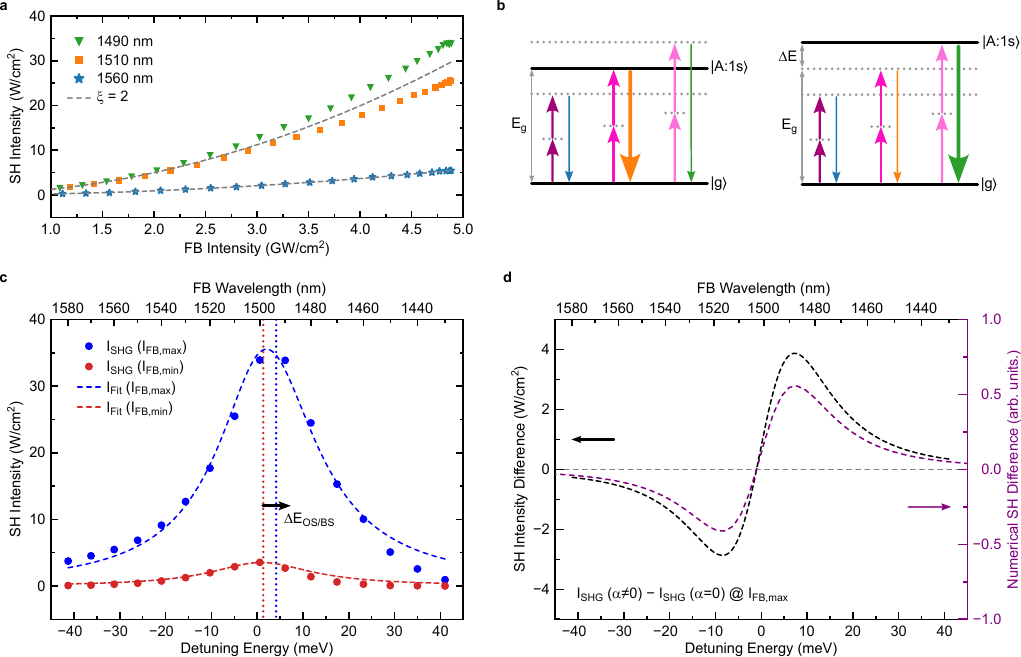}
    \caption{\label{fig:4}\textbf{Intensity-dependent SHG process and comparison with the theoretical model.} \textbf{a}, Power-dependent SHG datasets for \SIlist{1560;1510;1490}{\nano\meter} (blue, orange, and green dots) on a linear scale. The \qty{1560}{\nano\meter} data follows perfectly the lower gray dashed line, corresponding to a fit with $\xi = 2$. Instead, the scaling for \qty{1510}{} (\qty{1490}{\nano\meter}) decreases (increases) for higher power values compared to the upper gray dashed line. \textbf{b}, Schematic of scaling deviations: SHG emission is enhanced when two photons match the resonance condition with the ${A\!\!:\!\!1s}$ exciton state. A bandgap shift $\Delta E$ alters this resonance, increasing SHG efficiency for a different photon energy. \textbf{c}, Experimental second-harmonic intensities $I_{\text{SHG}}$ plotted for the approximate maximum ($I_{\text{FB,max}} \approx$\;\qty{5}{\giga\watt\per\centi\square\meter}; blue dots) and minimum FB intensity  ($I_{\text{FB,min}} \approx$\;\qty{1.5}{\giga\watt\per\centi\square\meter}, red dots). Dashed lines with same colors show the fitted intensities $I_{\text{Fit}}$ calculated with the analytical expression. Dotted lines indicate the calculated respective blueshift $\Delta E_{\text{OS/BS}}$ of the resonance by the OS and BS shifts for $E_{\text{det}}=0$\;\unit{\eV}. \textbf{d}, Comparison of the analytical SH intensity difference (black dashed line) caused by the perturbation $\alpha$ with our numerical simulations (purple dashed line).}
\end{figure}

Having established the method to filter the SH signal from the TP-PL background, we now focus on the SH power-dependent measurements in the spectral region of the ${A\!\!:\!\!1s}$ resonance. We analyze the power-dependent SHG data sets and their scaling (see Fig.\;4a) for FB wavelengths ranging from \qtyrange{1.43}{1.58}{\micro\meter}. In Fig.\;4a, we show a set of exemplary wavelengths, where we compare the scaling factor $\xi$, the proportionality between the generated SH signal, and the incident fundamental intensity ($I_\text{NLO}(\xi\omega)\propto I_\text{FB}^{\xi}(\omega)$), with a perfect quadratic scaling. Indeed, for a FB energy far below the resonance (\textit{e.g.} for a wavelength of \qty{1560}{\nano\meter}) we observe for the entire investigated power scan, up to $\sim$\;\qty{6}{\;\unit{\milli\watt}}, a scaling factor of $\xi = 2$, as expected for a second-order nonlinear process in the perturbative regime. However, when the FB wavelength is close to the resonance, we observe a significant deviation from the canonical value of 2, especially at large values of the FB power. In particular, the scaling factor $\xi$ is $<2$ for an excitation energy below the resonance (\textit{e.g.} at \qty{1510}{\nano\meter}) and $>2$ for an excitation energy above the resonance (\textit{e.g.} at \qty{1490}{\nano\meter}). As already discussed, the scaling factor $\xi$ is often considered the fingerprint of perturbative NLO, but our results clearly indicate the intrinsic limitations of this assumption. As we will discuss in the following, this clear deviation from $\xi = 2$ is due to the dual effect of light, which simultaneously modulates the bandgap by the intensity-dependent OS and BS shifts, and probes such modulation by resonant \textit{vs} non-resonant SHG, as schematically depicted in Fig.\;\ref{fig:4}b. It is worth noting that a deviation from a scaling factor $\xi = n$ for a $n$-th order NLO process has been reported for instance in the case of THG in graphene \cite{soavi2019hot} and SHG in metals \cite{papadogiannis1997nonlinear}. However, in both examples changes in the scaling factor $\xi$ are due to incoherent and time-dependent electron thermalization, and their observation is thus highly dependent on the excitation parameters, such as the pulse duration and repetition rate. In contrast, the mechanism seen here is fully coherent and independent of the presence of a photo-excited hot-electron distribution. 

In order to take the light-induced OS and BS effects into account, we introduce an intensity-dependent perturbation to the second-order susceptibility:
%
% \begin{align}
%     {|\chi_{yyy}|^2}
%     \label{eq:2}
%     &={|\chiint+\chi_{I}I|^2}=|\chiint|^2\left(1+2\frac{|\chi_{I}|}{|\chiint|}I\cos{\theta}+\frac{|\chi_{I}|^2}{|\chiint|^2}I^2\right)
% \end{align}
\begin{align}
 |\chi_{yyy}|^2 =\left |\chiint (1+\alpha I) \right|^2  
    = 
     |\chiint|^2 {
\left(1+2\text{Re}(\alpha)I  +|\alpha|^2I^2\right)}   \label{eq:2}
\end{align}
where $\chiint$ is the intrinsic (equilibrium) contribution, $\alpha$ a proportionality factor and $I$ the FB peak intensity. 
In our previous work~\cite{herrmann2025nonlinear}, we have developed an analytical model for the second-order susceptibility based on perturbative solutions of the SBEs. Our model incorporates the OS and BS shifts via a Floquet Hamiltonian and quantifies $\chiint$ and $ \alpha$ as (see Supplementary Section 2 for details)
\begin{align}
\chiint &= C_1\,\left[{\displaystyle 2E_{\text{det}}+\frac{i\hbar}{T_2}}\right]^{-1}\;,
\hspace{2em}
\alpha = C_2\,d^2
   \,\left[{\displaystyle 2E_{\text{det}}+\frac{i\hbar}{T_2}}\right]^{-1}\,.
  %I also replaced \Delta = \hbar c / (\lambda_\text{res]/2)}
  \label{eq:3}
\end{align}
Here, $C_1$ and  $C_2={8}/({3 c\epsilon n \Delta})$ are constants, $E_{\text{det}}={hc}/{\lambda_\text{FB}}-\Delta$ is the detuning energy from the resonance condition with the FB wavelength $\lambda_\text{FB}$ and half-gap energy $\Delta =$\;\qty{0.826}{\eV} (determined from the PL peak position in Fig. 2b). 
The unknown parameters include the dephasing time
$T_2$  and the absolute value of dipole moment at the $\pm$K-points, $d$.
To extract $T_2$ and $d$, we employ the SHG peak intensity~$I_\text{SHG}$~\cite{de2023nonlinear}
\begin{align}
    I_\text{SHG}
    \label{eq:4}
    &= \frac{C_3}{\lambda_\text{FB}^2}{|\chi_{yyy}|^2}I^2
\end{align}
with the real-valued constant $C_3 =$\;\qty{2.267e-15}{\square\volt\square\meter\per\watt}  (see details in the Supplementary Section~1).
We then combine equations (\ref{eq:2}), \eqref{eq:3} and (\ref{eq:4}), and fit the experimental SHG intensity $ I_\text{SHG}(\lambda_\text{FB},I)$ measured for various $\lambda_\text{FB}$ and $I$ to determine the parameters $d$ and $T_2$.
We obtain fitting values for the dipole moment $d = \qty{3.2}{} \pm \qty{0.2}{e\angstrom}$ and the dephasing time $T_2 =\qty{24}{} \pm \qty{1}{\femto\second}$, which align well with previously reported experimental values \cite{kim2014ultrafast,luo2023ultrafast}. 
It is important to note that the peak broadening observed in our experiments is affected by both homogeneous and inhomogeneous broadening mechanisms. Homogeneous broadening arises from processes such dephasing due to electron-electron and electron-phonon scattering, which are described by the dephasing time~$T_2$. Inhomogeneous broadening, on the other hand, results from local potential variations due to strain or defects \cite{moody2015intrinsic}. Our fitting  does not separate these different sources of broadening, and instead includes them all in an effective broadening value, $\hbar/T_2$, which is larger than the broadening only due to dephasing.  As a result, the fitted value $T_2 = \qty{24}{} \pm \qty{1}{\femto\second}$ should be considered as lower limit for the actual dephasing time.

In Fig.\;\ref{fig:4}c we show the measured second-harmonic intensity for the highest (blue dots) and lowest (red dots) evaluated FB intensities with respect to $E_{\text{det}}$ (and $\lambda_{\text{FB}}$), where we observe an increasing blueshift of the resonance depending on the FB intensity $I_{\text{FB}}$. This general trend is also nicely reproduced by the fitted analytical expression and the calculated shift $\Delta E_{\text{OS/BS}}$ for $E_{\text{det}}=0$\;\unit{\eV} \cite{herrmann2025nonlinear}.
%, where the maximum position of the curves (indicated by grey dashed lines) indicates the intensity dependent blueshift of the resonance, where we estimate a maximum shift $\Delta E_{\text{OS/BS}}=$\;\qty{4}{\milli\eV} for our experimental conditions and fitting parameters using the expression from Ref. \cite{herrmann2025nonlinear}.
To further verify our analytical model and the extracted fitting parameters, we compare the results with  numerical simulations, which use the same underlying Hamiltonian as the analytical model. In Fig.\;\ref{fig:4}d we show the SH intensity differences, comparing the SH values with and without the perturbation $\alpha$ from equation \ref{eq:2}, obtained from analytical calculations and numerical simulations at the largest FB intensity $I_{\text{FB,max}} =$\;\qty{5}{\giga\watt\per\centi\square\meter}, where the two approaches show an excellent qualitative agreement. We also extract from the numerical simulations another pair of values for the dipole moment $d^{\mathrm{sim}}=$\;\qty{4.9}{e\angstrom} and dephasing time $T_2^{\mathrm{sim}}= $\;\qty{25}{\femto\second} (see Supplementary Section 3 for details). 
While for the latter, the results from analytics and numerical simulations are in excellent agreement, the dipole element differs substantially. This discrepancy can be attributed to two distinct differences of the two theoretical approaches. First, the numerical simulations are run with pulsed excitation, whereas the analytical model assumes a continuous wave (CW) excitation. Since the average field of a pulse is smaller than in the CW case, the analytical function underestimates the dipole element involved. Second, the analytical model is based on approximations that are instead missing in the numerical simulations, which inherently takes into account all possible nonlinear effects that could affect the observed intensity-dependent blueshift of the SH signal.
%Second, in the analytical model we only include one contributing term of $\chi^{(4)}$ to the intensity-dependent blueshift of the second harmonic, whereas the numerical simulations inherently takes all contributions of $\chi^{(4)}$ into account. 
%Nevertheless, the combination of our two models provides a direct estimate of these fundamental material constants. Therefore, we determine the approximate dephasing time of $T_2=$\;\qty{25}{\femto\second} and a range for the dipole element from $d=$\;\qtyrange{3.2}{4.9}{e\angstrom}.
We thus determine the dipole element as $d =$\;\qty{4.9}{e\angstrom} and a lower limit of  $T_2 = $\;\qty{25}{\femto\second} for the dephasing time.

\subsection{Conclusion}
In this work, we investigated the intrinsic limitations and capabilities of parametric nonlinear optics, in particular as a spectroscopy tool to probe the equilibrium condition of resonant two-level systems. We have shown that, even in the perturbative regime, coherent all-optical modulation triggered by the fundamental beam can modify the energy levels under investigation. A deeper understanding of the dual effect of the light that simultaneously perturbs and measures the sample therefore plays an indispensable key role for any measurement of nonlinear optical effects in the perturbative regime. In our {WSe}$_2$ monolayer sample, this dual effect manifests itself in power-dependent measurements in the proximity of the A exciton resonance, with a characteristic deviation from the canonical quadratic scaling of the SH intensity. This unusual behavior is caused by an intensity-dependent blueshift of the energy levels caused by light-induced OS and BS shifts. We corroborate our experimental observations both analytically and numerically with a theoretical model based on the Semiconductor Bloch Equations. From this, we additionally extract the fundamental material parameters: dephasing time and transition dipole element, which can serve again as a basis to optimize the used models. Finally, we anticipate that SH modulation, achieved through optically controlled bandgap engineering in the perturbative regime, could enable novel nonlinear modulation schemes. When combined with elements like metasurfaces \cite{nauman2021tunable} or photonic cavities \cite{zhoucavity2024}, this approach could lead to sizeable modulations at even lower excitation intensities than demonstrated in this work. 

\section{Methods}
\subsection{Polarization resolved SHG}
For the power-dependent SHG measurements, we use the FB, generated by an optical parametric oscillator (Levante IR fs from APE), pumped by the output of a Yb doped mode-locked laser (FLINT FL2-12, Light Conversion) with a repetition rate of \qty{76}{\mega\hertz} and pulse length of \qty{100}{\femto\second}. This allows tuning of the FB from \qty{1300}{\nano\meter} to \qty{2000}{\nano\meter}. 

We guide the laser in a home-made multi photon microscopy setup, which we operate in transmission geometry. Before entering the microscope, a combination of halfwave-plate (AHWP05M-1600, Thorlabs) and wire-grid polarizer (WP25M-UB, Thorlabs) allows us to fully control the polarization axis of the FB. Subsequently, the FB is focussed onto the sample by a $\times 40$ objective (LMM-40X-P01, Thorlabs) and the transmitted FB, as well as the generated NLO signal are collimated by a lens (C330TMD, Thorlabs). The transmitted FB is blocked by a shortpass filter (FESH0950 \& FESH0850, Thorlabs) and the signal is further separated from the TH by an additional longpass filter (FEL0550, Thorlabs). With an additional Glan-Thompson prism (GTH10M, Thorlabs), we are able to selectively block the SH and transmit the TP-PL. Finally, we detect the remaining signal with a silicon avalanche-photo-diode (APD440A, Thorlabs) and lock-in amplifier (HF2LI, Zurich Instruments).

A similar setup was used to collect the 2D color maps in Fig.\;\ref{fig:3}a. However, for these, we used a Ti:Sapphire pump laser (Chameleon Ultra II) in combination with an optical parametric oscillator (Chameleon Compact OPO, both Coherent Inc.) for the FB. The output with a repetition rate of \qty{80}{\mega\hertz} and a pulse duration of \qty{220}{\femto\second} \cite{valencia2024enhanced} can be tuned from \qtyrange{1.1}{1.6}{\micro\meter}.

\subsection{Sample preparation and characterization}
We mechanically exfoliate {WSe}$_2$ from a bulk crystal (HQ Graphene) onto PDMS and transfer it onto a transparent fused silica substrate. To confirm the monolayer nature of the flake of interest, we evaluate PL and Raman measurements of the sample and ensure it shows the characteristic monolayer signatures. In particular, we look for a strong PL emission peak at the exciton resonance ($\sim$\;\qty{750}{\nano\meter} for WSe$_2$), the typical frequency difference between the degenerate {E$^{1}_{2g}$} and {A$_{1g}$} modes and the second-order {2LA(M)} mode, and the absence of the {B$^{1}_{2g}$} peak in the Raman spectrum. All these characteristics are present in the exfoliated sample (see Fig.\;\ref{fig:2}b,c), which confirms its monolayer nature. A home-built micro-photoluminescence spectroscopy system was used for the PL characterization. In the setup, a \SI{532}{\nano\meter} CW-laser was coupled into a commercial optical microscope (WiTec alpha300 S). The photoluminescence (PL) signal was collected in reflection geometry and guided to a grating spectrometer (QE-65000-FL, Ocean Optics) and evaluated with the internal detector. For the Raman spectroscopy, we used a commercial system (inVia Reflex, Renishaw) equipped with a \SI{532}{\nano\meter} CW-laser.

\section{Author Contributions Statement}

S.K., D.N., and G.S. conceived the experiments. S.K. and L.V.M. performed the nonlinear optical experiments and analyzed the data. S.K. prepared the monolayer sample by mechanical exfoliation. T.L. and U.P. performed the numerical simulations. J.W. developed the theoretical analytical model based on SBEs. S.K., J.W., U.P., D.N., and G.S. interpreted the experimental and theoretical results. S.K., T.L., J.W., and G.S. wrote the manuscript, with contributions from all authors. All authors participated in the discussion and commented on the manuscript.

\section{Acknowledgments}
We would like to thank Isabell Straller for her contribution to automating the experimental setup for the measurement of the polarization-resolved 2D color maps. This work was funded by the German Research Foundation DFG (CRC 1375 NOA), project number 398816777 (subproject C4); the International Research Training Group (IRTG) 2675 “Meta-Active”, project number 437527638 (subproject A4); and by the Federal Ministry for Education and Research (BMBF) project number 16KIS1792 SiNNER. J.W.~acknowledges the DFG for funding via the Emmy Noether Programme (project 503985532), CRC 1277 (project number 314695032, subproject A03) and RTG 2905 (project number~502572516). D.N. and L.V.M acknowledge the support of the Australian Research Council for funding via the CoE program (CE200100010). 

\section{Data availability}
The data that support the plots within this paper and other findings of this study are available from the corresponding author on reasonable request. Source data are provided with this paper.

%\bibliography{references}% Produces the bibliography via BibTeX.

\begin{thebibliography}{10}
\expandafter\ifx\csname url\endcsname\relax
  \def\url#1{\texttt{#1}}\fi
\expandafter\ifx\csname urlprefix\endcsname\relax\def\urlprefix{URL }\fi
\providecommand{\bibinfo}[2]{#2}
\providecommand{\eprint}[2][]{\url{#2}}

\bibitem{heisenberg1927anschaulichen}
\bibinfo{author}{Heisenberg, W.}
\newblock \bibinfo{title}{{\"U}ber den anschaulichen inhalt der quantentheoretischen kinematik und mechanik}.
\newblock \emph{\bibinfo{journal}{Zeitschrift f{\"u}r Physik}} \textbf{\bibinfo{volume}{43}}, \bibinfo{pages}{172--198} (\bibinfo{year}{1927}).

\bibitem{bohr1928quantum}
\bibinfo{author}{Bohr, N.}
\newblock \bibinfo{title}{The {Quantum} {Postulate} and the {Recent} {Development} of {Atomic} {Theory}}.
\newblock \emph{\bibinfo{journal}{Nature}} \textbf{\bibinfo{volume}{121}}, \bibinfo{pages}{580--590} (\bibinfo{year}{1928}).
\newblock \urlprefix\url{https://doi.org/10.1038/121580a0}.

\bibitem{heide2024petahertz}
\bibinfo{author}{Heide, C.}, \bibinfo{author}{Keathley, P.~D.} \& \bibinfo{author}{Kling, M.~F.}
\newblock \bibinfo{title}{Petahertz electronics}.
\newblock \emph{\bibinfo{journal}{Nature Reviews Physics}} \bibinfo{pages}{1--15} (\bibinfo{year}{2024}).

\bibitem{budden2021evidence}
\bibinfo{author}{Budden, M.} \emph{et~al.}
\newblock \bibinfo{title}{Evidence for metastable photo-induced superconductivity in k3c60}.
\newblock \emph{\bibinfo{journal}{Nature Physics}} \textbf{\bibinfo{volume}{17}}, \bibinfo{pages}{611--618} (\bibinfo{year}{2021}).

\bibitem{zeng2025photo}
\bibinfo{author}{Zeng, Z.} \emph{et~al.}
\newblock \bibinfo{title}{Photo-induced chirality in a nonchiral crystal}.
\newblock \emph{\bibinfo{journal}{Science}} \textbf{\bibinfo{volume}{387}}, \bibinfo{pages}{431--436} (\bibinfo{year}{2025}).

\bibitem{wang2013observation}
\bibinfo{author}{Wang, Y.}, \bibinfo{author}{Steinberg, H.}, \bibinfo{author}{Jarillo-Herrero, P.} \& \bibinfo{author}{Gedik, N.}
\newblock \bibinfo{title}{Observation of floquet-bloch states on the surface of a topological insulator}.
\newblock \emph{\bibinfo{journal}{Science}} \textbf{\bibinfo{volume}{342}}, \bibinfo{pages}{453--457} (\bibinfo{year}{2013}).

\bibitem{weitz2024lightwave}
\bibinfo{author}{Weitz, T.} \emph{et~al.}
\newblock \bibinfo{title}{Lightwave-driven electrons in a floquet topological insulator}.
\newblock \emph{\bibinfo{journal}{arXiv preprint arXiv:2407.17917}}  (\bibinfo{year}{2024}).

\bibitem{sie2018large}
\bibinfo{author}{Sie, E.~J.} \& \bibinfo{author}{Sie, E.~J.}
\newblock \bibinfo{title}{Large, valley-exclusive bloch-siegert shift in monolayer {WS$_2$}}.
\newblock \emph{\bibinfo{journal}{Coherent Light-Matter Interactions in Monolayer Transition-Metal Dichalcogenides}} \bibinfo{pages}{77--92} (\bibinfo{year}{2018}).

\bibitem{herrmann2025nonlinear}
\bibinfo{author}{Herrmann, P.} \emph{et~al.}
\newblock \bibinfo{title}{Nonlinear valley selection rules and all-optical probe of broken time-reversal symmetry in monolayer {WSe$_2$}}.
\newblock \emph{\bibinfo{journal}{Nature Photonics}} \textbf{\bibinfo{volume}{19}}, \bibinfo{pages}{300--306} (\bibinfo{year}{2025}).

\bibitem{maiuri2019ultrafast}
\bibinfo{author}{Maiuri, M.}, \bibinfo{author}{Garavelli, M.} \& \bibinfo{author}{Cerullo, G.}
\newblock \bibinfo{title}{Ultrafast spectroscopy: State of the art and open challenges}.
\newblock \emph{\bibinfo{journal}{Journal of the American Chemical Society}} \textbf{\bibinfo{volume}{142}}, \bibinfo{pages}{3--15} (\bibinfo{year}{2019}).

\bibitem{shree2021guide}
\bibinfo{author}{Shree, S.}, \bibinfo{author}{Paradisanos, I.}, \bibinfo{author}{Marie, X.}, \bibinfo{author}{Robert, C.} \& \bibinfo{author}{Urbaszek, B.}
\newblock \bibinfo{title}{Guide to optical spectroscopy of layered semiconductors}.
\newblock \emph{\bibinfo{journal}{Nature Reviews Physics}} \textbf{\bibinfo{volume}{3}}, \bibinfo{pages}{39--54} (\bibinfo{year}{2021}).

\bibitem{boyd2020nonlinear}
\bibinfo{author}{Boyd, R.~W.}
\newblock \emph{\bibinfo{title}{Nonlinear {Optics}}} (\bibinfo{publisher}{Academic Press}, \bibinfo{year}{2020}).

\bibitem{soavi2025signature}
\bibinfo{author}{Soavi, G.} \& \bibinfo{author}{Wilhelm, J.}
\newblock \bibinfo{title}{The signature of topology in polar and chiral non-magnetic crystal classes}.
\newblock \emph{\bibinfo{journal}{arXiv preprint arXiv:2501.03684}}  (\bibinfo{year}{2025}).

\bibitem{li2013probing}
\bibinfo{author}{Li, Y.} \emph{et~al.}
\newblock \bibinfo{title}{Probing symmetry properties of few-layer {MoS}$_2$ and {h-BN} by optical second-harmonic generation}.
\newblock \emph{\bibinfo{journal}{Nano letters}} \textbf{\bibinfo{volume}{13}}, \bibinfo{pages}{3329--3333} (\bibinfo{year}{2013}).

\bibitem{fiebig2023nonlinear}
\bibinfo{author}{Fiebig, M.}
\newblock \emph{\bibinfo{title}{Nonlinear optics on ferroic materials}} (\bibinfo{publisher}{John Wiley \& Sons}, \bibinfo{year}{2023}).

\bibitem{wang2015giant}
\bibinfo{author}{Wang, G.} \emph{et~al.}
\newblock \bibinfo{title}{Giant enhancement of the optical second-harmonic emission of {WS$_2$} monolayers by laser excitation at exciton resonances}.
\newblock \emph{\bibinfo{journal}{Physical review letters}} \textbf{\bibinfo{volume}{114}}, \bibinfo{pages}{097403} (\bibinfo{year}{2015}).

\bibitem{cunha2020second}
\bibinfo{author}{Cunha, R.} \emph{et~al.}
\newblock \bibinfo{title}{Second harmonic generation in defective hexagonal boron nitride}.
\newblock \emph{\bibinfo{journal}{Journal of Physics: Condensed Matter}} \textbf{\bibinfo{volume}{32}}, \bibinfo{pages}{19LT01} (\bibinfo{year}{2020}).

\bibitem{mennel2019second}
\bibinfo{author}{Mennel, L.}, \bibinfo{author}{Paur, M.} \& \bibinfo{author}{Mueller, T.}
\newblock \bibinfo{title}{Second harmonic generation in strained transition metal dichalcogenide monolayers: {MoS}$_2$, {MoSe}$_2$, {WS}$_2$, and {WSe}$_2$}.
\newblock \emph{\bibinfo{journal}{APL Photonics}} \textbf{\bibinfo{volume}{4}} (\bibinfo{year}{2019}).

\bibitem{ghimire2019high}
\bibinfo{author}{Ghimire, S.} \& \bibinfo{author}{Reis, D.~A.}
\newblock \bibinfo{title}{High-harmonic generation from solids}.
\newblock \emph{\bibinfo{journal}{Nature physics}} \textbf{\bibinfo{volume}{15}}, \bibinfo{pages}{10--16} (\bibinfo{year}{2019}).

\bibitem{zograf2022high}
\bibinfo{author}{Zograf, G.} \emph{et~al.}
\newblock \bibinfo{title}{High-harmonic generation from resonant dielectric metasurfaces empowered by bound states in the continuum}.
\newblock \emph{\bibinfo{journal}{Acs Photonics}} \textbf{\bibinfo{volume}{9}}, \bibinfo{pages}{567--574} (\bibinfo{year}{2022}).

\bibitem{SchmittRink1988}
\bibinfo{author}{Schmitt-Rink, S.}, \bibinfo{author}{Chemla, D.~S.} \& \bibinfo{author}{Haug, H.}
\newblock \bibinfo{title}{{Nonequilibrium theory of the optical Stark effect and spectral hole burning in semiconductors}}.
\newblock \emph{\bibinfo{journal}{Phys. Rev. B}} \textbf{\bibinfo{volume}{37}}, \bibinfo{pages}{941--955} (\bibinfo{year}{1988}).
\newblock \urlprefix\url{https://link.aps.org/doi/10.1103/PhysRevB.37.941}.

\bibitem{Lindberg1988}
\bibinfo{author}{Lindberg, M.} \& \bibinfo{author}{Koch, S.~W.}
\newblock \bibinfo{title}{{Effective Bloch equations for semiconductors}}.
\newblock \emph{\bibinfo{journal}{Phys. Rev. B}} \textbf{\bibinfo{volume}{38}}, \bibinfo{pages}{3342--3350} (\bibinfo{year}{1988}).
\newblock \urlprefix\url{https://link.aps.org/doi/10.1103/PhysRevB.38.3342}.

\bibitem{Wilhelm2020}
\bibinfo{author}{Wilhelm, J.} \emph{et~al.}
\newblock \bibinfo{title}{{Semiconductor Bloch-equations formalism: Derivation and application to high-harmonic generation from Dirac fermions}}.
\newblock \emph{\bibinfo{journal}{Phys. Rev. B}} \textbf{\bibinfo{volume}{103}}, \bibinfo{pages}{125419} (\bibinfo{year}{2021}).
\newblock \urlprefix\url{https://link.aps.org/doi/10.1103/PhysRevB.103.125419}.

\bibitem{Yue2022}
\bibinfo{author}{Yue, L.} \& \bibinfo{author}{Gaarde, M.~B.}
\newblock \bibinfo{title}{Introduction to theory of high-harmonic generation in solids: tutorial}.
\newblock \emph{\bibinfo{journal}{J. Opt. Soc. Am. B}} \textbf{\bibinfo{volume}{39}}, \bibinfo{pages}{535--555} (\bibinfo{year}{2022}).
\newblock \urlprefix\url{https://opg.optica.org/viewmedia.cfm?uri=josab-39-2-535&seq=0&html=true https://opg.optica.org/abstract.cfm?uri=josab-39-2-535 https://opg.optica.org/josab/abstract.cfm?uri=josab-39-2-535}.

\bibitem{Aversa1995}
\bibinfo{author}{Aversa, C.} \& \bibinfo{author}{Sipe, J.~E.}
\newblock \bibinfo{title}{Nonlinear optical susceptibilities of semiconductors: Results with a length-gauge analysis}.
\newblock \emph{\bibinfo{journal}{Phys. Rev. B}} \textbf{\bibinfo{volume}{52}}, \bibinfo{pages}{14636--14645} (\bibinfo{year}{1995}).
\newblock \urlprefix\url{https://link.aps.org/doi/10.1103/PhysRevB.52.14636}.

\bibitem{dogadov2022parametric}
\bibinfo{author}{Dogadov, O.}, \bibinfo{author}{Trovatello, C.}, \bibinfo{author}{Yao, B.}, \bibinfo{author}{Soavi, G.} \& \bibinfo{author}{Cerullo, G.}
\newblock \bibinfo{title}{Parametric nonlinear optics with layered materials and related heterostructures}.
\newblock \emph{\bibinfo{journal}{Laser \& Photonics Reviews}} \textbf{\bibinfo{volume}{16}}, \bibinfo{pages}{2100726} (\bibinfo{year}{2022}).

\bibitem{nauman2021tunable}
\bibinfo{author}{Nauman, M.} \emph{et~al.}
\newblock \bibinfo{title}{Tunable unidirectional nonlinear emission from transition-metal-dichalcogenide metasurfaces}.
\newblock \emph{\bibinfo{journal}{Nature communications}} \textbf{\bibinfo{volume}{12}}, \bibinfo{pages}{5597} (\bibinfo{year}{2021}).

\bibitem{wang2018colloquium}
\bibinfo{author}{Wang, G.} \emph{et~al.}
\newblock \bibinfo{title}{Colloquium: Excitons in atomically thin transition metal dichalcogenides}.
\newblock \emph{\bibinfo{journal}{Reviews of Modern Physics}} \textbf{\bibinfo{volume}{90}}, \bibinfo{pages}{021001} (\bibinfo{year}{2018}).

\bibitem{Klimmer2021all-optical}
\bibinfo{author}{Klimmer, S.} \emph{et~al.}
\newblock \bibinfo{title}{All-optical polarization and amplitude modulation of second-harmonic generation in atomically thin semiconductors}.
\newblock \emph{\bibinfo{journal}{Nature Photonics}} \textbf{\bibinfo{volume}{15}}, \bibinfo{pages}{837--842} (\bibinfo{year}{2021}).
\newblock \urlprefix\url{https://doi.org/10.1038/s41566-021-00859-y}.

\bibitem{trovatello2021optical}
\bibinfo{author}{Trovatello, C.} \emph{et~al.}
\newblock \bibinfo{title}{Optical parametric amplification by monolayer transition metal dichalcogenides}.
\newblock \emph{\bibinfo{journal}{Nature Photonics}} \textbf{\bibinfo{volume}{15}}, \bibinfo{pages}{6--10} (\bibinfo{year}{2021}).

\bibitem{trovatello2024tunable}
\bibinfo{author}{Trovatello, C.} \emph{et~al.}
\newblock \bibinfo{title}{Tunable optical nonlinearities in layered materials}.
\newblock \emph{\bibinfo{journal}{ACS Photonics}} \textbf{\bibinfo{volume}{11}}, \bibinfo{pages}{2860--2873} (\bibinfo{year}{2024}).

\bibitem{trovatello2025quasi}
\bibinfo{author}{Trovatello, C.} \emph{et~al.}
\newblock \bibinfo{title}{Quasi-phase-matched up-and down-conversion in periodically poled layered semiconductors}.
\newblock \emph{\bibinfo{journal}{Nature Photonics}} \bibinfo{pages}{1--9} (\bibinfo{year}{2025}).

\bibitem{herrmann2023nonlinear}
\bibinfo{author}{Herrmann, P.} \emph{et~al.}
\newblock \bibinfo{title}{Nonlinear all-optical coherent generation and read-out of valleys in atomically thin semiconductors}.
\newblock \emph{\bibinfo{journal}{Small}} \textbf{\bibinfo{volume}{19}}, \bibinfo{pages}{2301126} (\bibinfo{year}{2023}).

\bibitem{conway2023effects}
\bibinfo{author}{Conway, M.~A.} \emph{et~al.}
\newblock \bibinfo{title}{Effects of floquet engineering on the coherent exciton dynamics in monolayer {WS$_2$}}.
\newblock \emph{\bibinfo{journal}{ACS nano}} \textbf{\bibinfo{volume}{17}}, \bibinfo{pages}{14545--14554} (\bibinfo{year}{2023}).

\bibitem{zhoucavity2024}
\bibinfo{author}{Zhou, L.} \emph{et~al.}
\newblock \bibinfo{title}{Cavity floquet engineering}.
\newblock \emph{\bibinfo{journal}{Nature Communications}} \textbf{\bibinfo{volume}{15}}, \bibinfo{pages}{7782}.
\newblock \urlprefix\url{https://doi.org/10.1038/s41467-024-52014-0}.

\bibitem{gucci2024ultrafast}
\bibinfo{author}{Gucci, F.} \emph{et~al.}
\newblock \bibinfo{title}{Ultrafast valleytronic logic operations}.
\newblock \emph{\bibinfo{journal}{arXiv preprint arXiv:2412.08318}}  (\bibinfo{year}{2024}).

\bibitem{rosa2018characterization}
\bibinfo{author}{Rosa, H.~G.} \emph{et~al.}
\newblock \bibinfo{title}{Characterization of the second-and third-harmonic optical susceptibilities of atomically thin tungsten diselenide}.
\newblock \emph{\bibinfo{journal}{Scientific reports}} \textbf{\bibinfo{volume}{8}}, \bibinfo{pages}{10035} (\bibinfo{year}{2018}).

\bibitem{lafeta2021second}
\bibinfo{author}{Lafeta, L.} \emph{et~al.}
\newblock \bibinfo{title}{Second-and third-order optical susceptibilities across excitons states in {2D} monolayer transition metal dichalcogenides}.
\newblock \emph{\bibinfo{journal}{{2D} Materials}} \textbf{\bibinfo{volume}{8}}, \bibinfo{pages}{035010} (\bibinfo{year}{2021}).

\bibitem{hernandez2021nonlinear}
\bibinfo{author}{Hernandez-Rueda, J.}, \bibinfo{author}{Noordam, M.~L.}, \bibinfo{author}{Komen, I.} \& \bibinfo{author}{Kuipers, L.}
\newblock \bibinfo{title}{Nonlinear optical response of a {WS$_2$} monolayer at room temperature upon multicolor laser excitation}.
\newblock \emph{\bibinfo{journal}{ACS photonics}} \textbf{\bibinfo{volume}{8}}, \bibinfo{pages}{550--556} (\bibinfo{year}{2021}).

\bibitem{saynatjoki2017ultra}
\bibinfo{author}{S{\"a}yn{\"a}tjoki, A.} \emph{et~al.}
\newblock \bibinfo{title}{Ultra-strong nonlinear optical processes and trigonal warping in {MoS$_2$} layers}.
\newblock \emph{\bibinfo{journal}{Nature communications}} \textbf{\bibinfo{volume}{8}}, \bibinfo{pages}{893} (\bibinfo{year}{2017}).

\bibitem{valencia2024enhanced}
\bibinfo{author}{Valencia~Molina, L.} \emph{et~al.}
\newblock \bibinfo{title}{Enhanced infrared vision by nonlinear up-conversion in nonlocal metasurfaces}.
\newblock \emph{\bibinfo{journal}{Advanced Materials}} \textbf{\bibinfo{volume}{36}}, \bibinfo{pages}{2402777} (\bibinfo{year}{2024}).

\bibitem{lin2020fast}
\bibinfo{author}{Lin, K.-I.} \emph{et~al.}
\newblock \bibinfo{title}{Fast real-space imaging of the exciton complexes in {WSe}$_2$ and {WS}$_2$ monolayers using multiphoton microscopy}.
\newblock \emph{\bibinfo{journal}{The Journal of Physical Chemistry C}} \textbf{\bibinfo{volume}{124}}, \bibinfo{pages}{7979--7987} (\bibinfo{year}{2020}).

\bibitem{glazov2017intrinsic}
\bibinfo{author}{Glazov, M.~M.} \emph{et~al.}
\newblock \bibinfo{title}{Intrinsic exciton-state mixing and nonlinear optical properties in transition metal dichalcogenide monolayers}.
\newblock \emph{\bibinfo{journal}{Physical Review B}} \textbf{\bibinfo{volume}{95}}, \bibinfo{pages}{035311} (\bibinfo{year}{2017}).

\bibitem{lundt2019optical}
\bibinfo{author}{Lundt, N.} \emph{et~al.}
\newblock \bibinfo{title}{Optical valley hall effect for highly valley-coherent exciton-polaritons in an atomically thin semiconductor}.
\newblock \emph{\bibinfo{journal}{Nature nanotechnology}} \textbf{\bibinfo{volume}{14}}, \bibinfo{pages}{770--775} (\bibinfo{year}{2019}).

\bibitem{soavi2019hot}
\bibinfo{author}{Soavi, G.} \emph{et~al.}
\newblock \bibinfo{title}{Hot electrons modulation of third-harmonic generation in graphene}.
\newblock \emph{\bibinfo{journal}{ACS Photonics}} \textbf{\bibinfo{volume}{6}}, \bibinfo{pages}{2841--2849} (\bibinfo{year}{2019}).

\bibitem{papadogiannis1997nonlinear}
\bibinfo{author}{Papadogiannis, N.} \& \bibinfo{author}{Moustaizis, S.}
\newblock \bibinfo{title}{Nonlinear enhancement of the efficiency of the second harmonic radiation produced by ultrashort laser pulses on a gold surface}.
\newblock \emph{\bibinfo{journal}{Optics communications}} \textbf{\bibinfo{volume}{137}}, \bibinfo{pages}{174--180} (\bibinfo{year}{1997}).

\bibitem{de2023nonlinear}
\bibinfo{author}{De~Matos, C.~J.} \emph{et~al.}
\newblock \bibinfo{title}{Nonlinear optics in {2D} materials: focus on the contributions from latin america}.
\newblock \emph{\bibinfo{journal}{Journal of the Optical Society of America B}} \textbf{\bibinfo{volume}{40}}, \bibinfo{pages}{C111--C132} (\bibinfo{year}{2023}).

\bibitem{kim2014ultrafast}
\bibinfo{author}{Kim, J.} \emph{et~al.}
\newblock \bibinfo{title}{Ultrafast generation of pseudo-magnetic field for valley excitons in {WSe$_2$} monolayers}.
\newblock \emph{\bibinfo{journal}{Science}} \textbf{\bibinfo{volume}{346}}, \bibinfo{pages}{1205--1208} (\bibinfo{year}{2014}).

\bibitem{luo2023ultrafast}
\bibinfo{author}{Luo, W.} \emph{et~al.}
\newblock \bibinfo{title}{Ultrafast nanoimaging of electronic coherence of monolayer {WSe}$_2$}.
\newblock \emph{\bibinfo{journal}{Nano Letters}} \textbf{\bibinfo{volume}{23}}, \bibinfo{pages}{1767--1773} (\bibinfo{year}{2023}).

\bibitem{moody2015intrinsic}
\bibinfo{author}{Moody, G.} \emph{et~al.}
\newblock \bibinfo{title}{Intrinsic homogeneous linewidth and broadening mechanisms of excitons in monolayer transition metal dichalcogenides}.
\newblock \emph{\bibinfo{journal}{Nature communications}} \textbf{\bibinfo{volume}{6}}, \bibinfo{pages}{8315} (\bibinfo{year}{2015}).

\end{thebibliography}

\end{document}

%\preprint{APS/123-QED}

\section{Title}
Supplementary Information --- Ultrafast Coherent Bandgap Modulation Probed by Parametric Nonlinear Optics

\section{Author list}
Sebastian Klimmer$^{1,2}$, Thomas Lettau$^3$, Laura Valencia Molina$^{1,2}$, Daniil Kartashov$^{4,5}$, Ulf Peschel$^{3,5}$, Jan Wilhelm$^{6,7}$, Dragomir Neshev$^{2}$ and Giancarlo Soavi$^{1,5,\star}$

\section{Affiliations}
\noindent
$^1$Institute of Solid State Physics, Friedrich Schiller University Jena, Helmholtzweg 5, 07743 Jena, Germany
\newline
$^2$ARC Centre of Excellence for Transformative Meta-Optical Systems, Department of Electronic Materials Engineering, Research School of Physics, The Australian National University, Canberra, ACT, 2601, Australia
\newline
$^3$Institute of Condensed Matter Theory and Optics, Friedrich Schiller University Jena, Max-Wien-Platz 1, 07743 Jena, Germany
\newline
$^4$Institute of Optics and Quantum Electronics, Friedrich Schiller University Jena, Max-Wien-Platz 1, 07743 Jena, Germany
\newline
$^5$Abbe Center of Photonics, Friedrich Schiller University Jena, Albert-Einstein-Straße 6, 07745 Jena, Germany
\newline
$^6$Regensburg Center for Ultrafast Nanoscopy (RUN), University of Regensburg, 93040 Regensburg, Germany
\newline
$^7$Institute of Theoretical Physics, University of Regensburg, 93053 Regensburg, Germany
\newline

$^{\star}$ giancarlo.soavi@uni-jena.de

%\keywords{nonlinear optics, SHG, valleytronics, 2d materials}
\maketitle

\subsection{Sheet Polarization Formalism for Second-Harmonic Generation}
In order to describe the NLO frequency conversion of 2D materials, we follow the widely accepted approach of calculating the NLO sheet susceptibility \cite{shen1989optical}. Subsequently, the effective bulk-like susceptibility 
can be obtained by multiplication with the film thickness.

Here, we reiterate briefly the derivation from Ref. \cite{de2023nonlinear}, with small changes to account for the used setup geometry and pulsed excitation, to obtain an expression for the second-order sheet susceptibility and to directly connect the fundamental and second-harmonic average powers.
\begin{figure}[h!]
    \centering
    \includegraphics{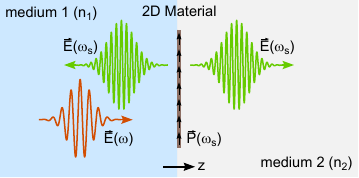}
    \caption{\label{fig:S1}\textbf{Schematic for the Nonlinear Sheet Polarization} A pump pulse (red) with frequency $\omega$ induces a nonlinear polarization $\boldsymbol{P}(\omega_S)$ (black arrows) in a 2D material (brown) at the interface of medium 1 and 2 with refractive indices $n_{1/2}$. The nonlinear sheet radiates the nonlinear signal (green) at frequency $\omega_S$ in both media.}
\end{figure}
Initially we consider the situation where a 2D polarization sheet (\textit{e.g.} a TMD monolayer) lies at the interface (at z=0) between two optically linear and isotropic bulk media $k$ ($k = 1,2$) with refractive indices $n_k$. The pump beam with frequency $\omega$, under normal incidence from medium 1, induces a nonlinear polarization $\boldsymbol{P}(\omega_S)$, which as a consequence radiates new electric fields $\boldsymbol{E}_{1,2}$ with the frequency $\omega_S$ into both media:
\begin{align}
    \boldsymbol{E}_{1,2}(\omega_S) = \frac{i\omega_S}{2c\epsilon_0(n_1+n_2)}[P_x(\omega_S) \hat{x}+P_y(\omega_S) \hat{y}],
\end{align}

where $\epsilon_0$ is the vacuum permittivity, and $c$ is the speed of light in vacuum. For second-harmonic generation we can redefine the emitted frequency $\omega_S = 2\omega$ and specifically write the nonlinear sheet polarization as:
\begin{align}
    \boldsymbol{P}^{(2)}_{SHG}(2\omega) = \epsilon_0 \boldsymbol{\chi}^{(2)}_S(2\omega,\omega,\omega)\boldsymbol{E}_{Sheet}^2(\omega),
\end{align}
where $\boldsymbol{\chi}^{(2)}_S(2\omega,\omega,\omega)$ is the nonlinear sheet susceptibility and $\boldsymbol{E}_{Sheet}$ is the pump field coupled into the polarization sheet, which is scaled by a factor $L(\omega)=2n_1(\omega)/(n_1(\omega)+n_2(\omega))$, derived from the Fresnel equations.
Now, we can write the electric field of the SHG signal, out-coupled into medium 2, as:
\begin{align}
    E_{2,SHG}(2\omega) = \frac{i2\omega L^2(\omega)L(2\omega)}{4cn_2(2\omega)}\boldsymbol{\chi}^{(2)}_S(2\omega,\omega,\omega)\boldsymbol{\hat{e}}(\omega)\boldsymbol{\hat{e}}(\omega)E_{1}^2(\omega),
\end{align}
where $\boldsymbol{E}_{1}(\omega) = E_{1}(\omega)\boldsymbol{\hat{e}}(\omega)$ is the incident pump field with the unit vector $\boldsymbol{\hat{e}}(\omega)$ indicating the pump polarization. 
In the following step we replace the electric fields with the peak intensities \textit{via} the relation $I_{k,peak}(\omega)=\frac{1}{2}\epsilon_0n_kc|E_k(\omega)|^2$:
\begin{gather}
\begin{aligned}
    \left|E^2_{2,SHG}(2\omega)\right| &= \left|\frac{i\omega L^2(\omega)L(2\omega)}{2cn_2(2\omega)}\boldsymbol{\chi}^{(2)}_SE_{1}^2(\omega)\right|^2,\\
    \frac{2I_{2,SHG}(2\omega)}{\epsilon_0n_2(2\omega)c} &= \frac{\omega^2 L^4(\omega)L^2(2\omega)}{4c^2n^2_2(2\omega)}|\chi^{(2)}_S|^2\left(\frac{2I_{1}(\omega)}{\epsilon_0n_1(\omega)c}\right)^2,\\
    I_{2,SHG}(2\omega) &= \frac{\omega^2L^4(\omega)L^2(2\omega)}{2\epsilon_0c^3n_2(2\omega)n_1^2(\omega)}|\chi^{(2)}_S|^2I_{1}(\omega)^2,
\end{aligned}
\end{gather}
We further assume that the 2D material is placed on a fused silica substrate with low dispersion (\textit{i.e.} $n_1(\omega) = n \approx 1.45$) and the SHG is emitted in air ($n_2 = 1$):
\begin{align}
    I_{2,SHG}(2\omega) &= \frac{32\omega^2n^2}{\epsilon_0c^3(1+n)^6}|\chi^{(2)}_S|^2I_{1}(\omega)^2
\end{align}
In a final step we rewrite the equation in terms of the fundamental wavelength $\lambda_{FB}$ instead of the frequency $\omega$ and replace the nonlinear sheet susceptibility with the product of the monolayer thickness $d_{\text{WSe$_2$}}\approx$\;\qty{0.7}{\nano\meter} \cite{seyler2015electrical} and the bulk-like effective susceptibility $\chi^{(2)}$:
\begin{equation}
\begin{aligned}
      I_{2,SHG}(\lambda_{\text{FB}}) &= \frac{128\pi^2n^2d_{\text{WSe$_2$}}^2}{\epsilon_0c(1+n)^6\lambda_{\text{FB}}^2}|\chi^{(2)}|^2I_{1}(\lambda_{\text{FB}})^2,\\
      &=\frac{C_3}{\lambda_\text{FB}^2}{|\chi^{(2)}|^2}I_{1}(\lambda_{\text{FB}})^2,
\end{aligned}
\end{equation}
combining all constant quantities in the real-valued constant $C_3 \approx$\;\qty{2.267e-15}{\square\volt\square\meter\per\watt}.

In order to evaluate this equation with our measured average powers $P(\lambda_{\text{FB}})$ of the FB pump and the generated SHG, we further convert them to peak intensities with the expression:
\begin{align}
    I_{peak}(\lambda_{\text{FB}})=\frac{2P(\lambda_{\text{FB}})S}{\pi w^2 f t},
\end{align}
with the Gaussian shape parameter $S=\sqrt{\frac{4\ln{(2)}}{\pi}}$, the repetition rate $f$, the 1/e$^2$ focal radius $w$ and the full-width at half-maximum pulse duration $t$. For the SHG intensity, we further account for a reduction of the pulse duration and focal radius by a factor of $\sqrt{2}$.

\subsection{Numerical methods}
\label{sec:numerical_methods}
To verify the validity of the analytical expression introduced in eq.~(4) of the main document, we numerically solved the Semiconductor Bloch Equations (SBEs) \cite{Lindberg1988,Wilhelm2020,Yue2022}
\begin{align}
    \label{eq:sbe}
    i\pdiff{t} \rho_{mn}(\bk;t) =& \left[\epsilon_m(\bkt) - \epsilon_n(\bkt) -(1-\delta_{nm})\frac{i}{T_2} \right]\rho_{mn}(\bk;t) \nonumber\\
     &+ \bE(t) \cdot \sum_{l} \left[\bd_{ml}(\bkt)\rho_{ln}(\bk;t) - \rho_{ml}(\bk;t)\bd_{ln}(\bkt)\right],
\end{align}
where $\rho_{mn}(\bk;t)$ are density matrix elements, $\epsilon_m(\bk)$ are the band energies for band $m$ and $\bd_{mn}(\bk) = i\Braket{u^{\bk}_{m}|\nabla_{\bk}u^{\bk}_{n}}$ are the dipole matrix elements (including the Berry connections for $n=m$) calculated from the lattice-periodic part of a Bloch function $u_{n\bk}$. $\bE$ is the electric field, $\bA=\int_{-\infty}^{t}\bE(t')\diff{t'}$ the associated vector potential, $\bkt=\bk-\bA(t)$ is a shifted $k$ vector, and $T_2$ is the dephasing time. Throughout this work, we use the convention $\hbar=1$. Analogously to our previous work \cite{herrmann2025nonlinear}, in which we describe the numerical solution of the SBEs in detail, we used a Haldane model \cite{Taghizadeh2019} to calculate the band energies and dipole matrix elements for eq.~(\ref{eq:sbe}). The electric field is parametrized by 
\begin{equation}
    \bE(t) = \Re\left[\hat{e}_{x}\, E_0 \exp^2\left(-t^{2}\frac{\log(4)^{2}}{T^{2}}\right)\exp(-i\omega_0 t)\right]\,, \label{efield}
\end{equation}
where $\hat{e}_{x}$ is the unit vector in $x$-direction, $E_0$ is the field strength and $T=200$fs is the FWHM of the Gaussian envelope with respect to the intensity of the light field. The frequency is varied around the half band gap $\Delta=\qty{0.826}{\eV}$ in 96 steps from $0.9\Delta$ to $1.1\Delta$. For each frequency, we simulate the SBEs for 20 different powers from $\qty{0.18}{\milli\watt}$ to $\qty{6.0}{\milli\watt}$, which correspond to field strengths of $\qty{0.025}{\volt\per\nano\meter}$ to $\qty{0.19}{\volt\per\nano\meter}$.
% TODO: explain how to convert from power to field strength
After simulating the SBEs, we calculate the induced current density
\begin{align}
     \bj(t) = -\frac{1}{N}\sum_{\bk\in\{-K,+K\}}\bj(\bk,t)\;, 
     \hspace{2em}
     \bj(\bk,t) := \sum_{n,m}\bp_{mn}(\bkt) \,\rho_{nm}(\bk;t),
       \label{eq:j}
\end{align}
where $N$ refers to the number of $k$-points in the Brillouin zone and $\bp_{mn}(\bk) = \nabla_{\bk}\epsilon_n(\bk)\delta_{mn} + i(\epsilon_m(\bk) - \epsilon_n(\bk))\bd_{mn}(\bk)$ are the momentum matrix elements. The sum over the $\bk$-points only contains one point at each valley $-K$ and $+K$. As the time derivative of the polarization is equal to the current density $\bj(t) = \pdiff{t}\bP(t)$, the susceptibility $\chi_{xxx}(\omega)$ can be determined by Fourier transforming the current density
\begin{equation}
    \chi_{xxx}(\omega) = \frac{i\omega j_{x}(\omega)}{E_{x}(\omega)^{2}},
\end{equation}
where $E_{x}(\omega)$ and $j_{x}(\omega)$ are the $x$-components of $\bE(\omega)$ and $\bj(\omega)$, respectively.

\subsection{Relating the analytical to the simulation Parameters}
To derive the analytical expression used in the present work (eq.~(4) of the main document), some approximations had to be made \cite{herrmann2025nonlinear}, in particular, the envelope of the pulse was not taken into account. As a result, the dephasing time $T_2^{\mathrm{ana}}$ fully determines the spectral width of the second-order susceptibility. However, since the finite pulse length should also influence the spectral width, we expect the dephasing time used in the simulation $T_2^{\mathrm{sim}}$ to be slightly larger. 
Similarly, the perturbation of the second-order susceptibility scales with the peak intensity of the field and the dipole element $d^{\mathrm{ana}}$. For a pulsed excitation, the proportionality factor is smaller, which results in a larger dipole element $d^{\mathrm{sim}}$ in the simulation.
\begin{figure}[h!]
    \centering
        \includegraphics[width=\textwidth]{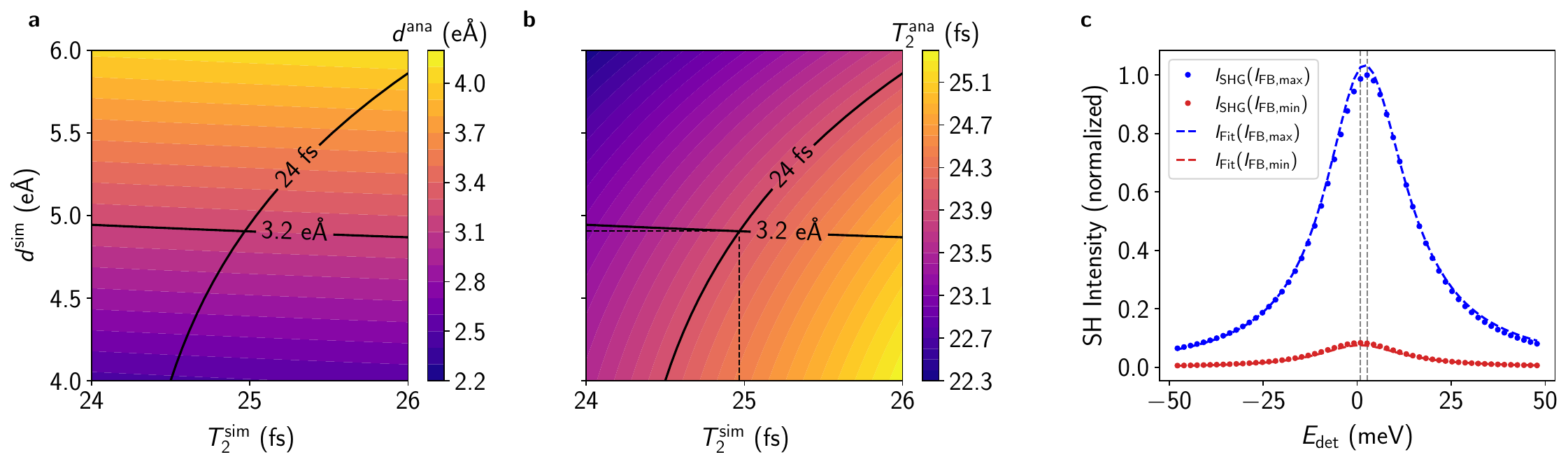}
    \caption{\label{fig:methE}\textbf{Determining the optimal simulation parameters.} 
    % TODO: Rename chi(0) to chi_int
    % TODO: Colormap is reversed
    % TODO: change T_2opt to T_2sim
    a) Fitted analytical parameter $d^{\mathrm{ana}}$ and b) $T_2^{\mathrm{ana}}$ as a function of the simulation parameters $d^{\mathrm{sim}}$ and $T_2^{\mathrm{sim}}$. The black contour lines show the analytical parameters determined from the experimental data. b) The intersection of the contour lines corresponds to the best simulation parameters. c) Analogously to Fig.~4c of the main document, the normalized simulated second-harmonic intensities as a function of the detuning energy $E_{\mathrm{det}}$ for the maximum (blue dots) and minimum FB intensity (red dots). The dashed lines of the same color indicate the fit to the analytical model, the gray dashed lines show the maxima of the fitted lines.}
\end{figure}
To find the corresponding parameters, we run the numerical simulations for $T_2^{\mathrm{sim}}\in[24, 26]\unit{\femto\second}$ and $d^{\mathrm{sim}} \in [4, 6]\unit{\elementarycharge\angstrom}$ in 21 steps each. Subsequently, we fitted the calculated second harmonic intensity to the analytical model, which relates the pair of simulation parameters ($T_2^{\mathrm{sim}}$, $d^{\mathrm{sim}}$) to a pair of analytical parameters ($T_2^{\mathrm{ana}}$, $d^{\mathrm{ana}}$). In Fig.~\ref{fig:methE}, we show the analytical parameters as a function of the simulation parameters in panels a) and b). In the same panels, we depict the analytical values that we extracted from the experiment as black contour lines at $T_2^{\mathrm{ana}}=\qty{24}{fs}$ and $d^{\mathrm{ana}}=\qty{3.2}{\elementarycharge\angstrom}$. Their intersection at $T_2^{\mathrm{sim}}\simeq\qty{25}{fs}$ and $d^{\mathrm{sim}}\simeq\qty{4.9}{\elementarycharge\angstrom}$ corresponds to the set of simulation parameters that lead to the same fit as the experimental results. 
%%%
%
% Keep for possible question from reviewer
%%While we observe a good qualitative agreement in the width and the general shape of the two components, we observe a deviation from the experimental data for positive detuning values. We attribute this to the influence of residual TP-PL in the experimental data. Since the SHG efficiency quickly drops and the TP-PL intensity simultaneously rises at the high-energy side of the resonance, errors in the SHG filtering with the non-perfect achromatic optical elements are here more pronounced, which leads to a slight underestimation of the intrinsic and as a consequence a simultaneous overestimation of the intensity-dependent susceptibility contribution.

%\bibliography{references}% Produces the bibliography via BibTeX.